\documentclass{aa}  %%%% DO NOT DELETE [oldversion]
\usepackage{graphics}
\usepackage{txfonts}
\usepackage{longtable}

\begin{document}

\title{HST and VLT observations of the Symbiotic Star Hen~2--147\thanks{Based on observations with the NASA/ESA Hubble Space
Telescope, obtained at the Space Telescope Science Institute, which is
operated by the Association of Universities for Research in Astronomy,
Inc. under NASA contract No.\ NAS5-26555; on
observations obtained at the 8m~VLT telescope of
the European Southern Observatory in Chile; and on observations made at the South African Astronomical Observatory}}

\subtitle{Its Nebular Dynamics, its Mira Variable and its Distance}

\author{M. Santander-Garc\'\i a\inst{1}
        \and R. L. M. Corradi\inst{2,1} 
        \and P. A. Whitelock\inst{3,4} 
        \and U. Munari\inst{5}
        \and A. Mampaso\inst{1}
        \and F. Marang\inst{3}
        \and F. Boffi\inst{6}
	\and M. Livio\inst{6}        }

\institute{
         Instituto de Astrof\'\i sica de Canarias, 38200 La Laguna,
        Tenerife, Spain
         \\e-mail: miguelsg@iac.es; amr@iac.es
        \and
         Isaac Newton Group of Telescopes, Ap.\ de Correos 321,
         38700 Sta. Cruz de la Palma, Spain
         \\e-mail: rcorradi@ing.iac.es
        \and
        South African Astronomical Observatory, PO Box 9, 7935 Observatory,
        South Africa
        \\e-mail: paw@saao.ac.za; fm@saao.ac.za
        \and
        Department of Mathematics and Applied Mathematics and Department of Astronomy,
        University of Cape Town, South Africa
        \and
        INAF Osservatorio Astronomico di Padova, via dell'Osservatorio 8,
        36012 Asiago (VI), Italy
        \\e-mail:munari@pd.astro.it
        \and
        Space Telescope Science Institute (STScI), 3700 San Martin Drive,
         Baltimore, MD 21218, USA
        e-mail: boffi@stsci.edu, mlivio@stsci.edu
        }

\offprints{M. Santander-Garc\'\i a}

\date{\today}
 
\abstract{}{We investigate the dynamics of the nebula around the
symbiotic star Hen~2--147, determine its expansion parallax, and
compare it with the distance obtained via the Period-Luminosity
relation for its Mira variable.}{A combination of multi-epoch HST
images and VLT integral field high-resolution spectroscopy is used to
study the nebular dynamics both along the line of sight and in the
plane of the sky. These observations allow us to build a 3-D
spatio--kinematical model of the nebula, which together with the
measurement of its apparent expansion in the plane of the sky over a
period of 3 years, provides the expansion parallax for the
nebula. Additionally, SAAO near-infrared photometry obtained over 25
years is used to determine the Mira pulsation period and derive an
independent distance estimation via the Period-Luminosity relationship
for Mira variables.}{The geometry of the nebula is found to be that of
a knotty annulus of ionized gas inclined to the plane of sky and
expanding with a velocity of $\sim90$ km~s$^{-1}$. A straightforward
application of the expansion parallax method provides a distance of
1.5$\pm$0.4~kpc, which is a factor of two lower than the
distance of $3.0 \pm 0.4$~kpc obtained from the Period-Luminosity
relationship for the Mira (which has a pulsation period of 373 days).
The discrepancy is removed if, instead of expanding matter, we are
observing the expansion of a shock front in the plane of the sky. This
shock interpretation is further supported by the broadening of the
nebular emission lines.}{}


\authorrunning{Santander-Garc\'\i a et al.}
\titlerunning{The symbiotic star Hen~2--147}
\maketitle

\section{Introduction}

The study of nebulae around symbiotic stars, and in particular those
containing Mira variables, is key to understanding several aspects of
their binary interactions, outburst properties and timescales. Such
studies provide information about the geometry, dynamics, and history
of mass loss from these systems over the last few thousand years.  It
is generally believed that in these binaries some percent of the Mira
wind is accreted by the white dwarf companion (Kenyon \& Webbink
\cite{KW84}), resulting in stable nuclear H-burning at the surface of
the white dwarf for some systems (e.g. Sokoloski \cite{So02}), and
sometimes causing thermonuclear outbursts (e.g. Kenyon \cite{Ke86},
Livio et al. \cite{Li89}) lasting for hundreds of years (e.g. Munari
\cite{Mu97}). The ejecta and fast wind produced during and after such
an outburst collide with the slowly expanding circum-binary gas
resulting from the non-accreted Mira wind.  This results in a complex
circum-binary nebula whose ionization mechanism (radiation and/or
shocks) is not well understood (Corradi \& Schwarz \cite{Co97}).

Currently, about a dozen symbiotic nebulae have been optically resolved
(Corradi \cite{Co03}). Some of them exhibit spectacular (mainly bipolar)
morphologies, such as R Aqr (Solf \& Ulrich \cite{So85}, Paresce \& Hack
\cite{Pa94} and references therein), Hen~2-104 and BI Cru (Schwarz et al.
\cite{Sw89}, Corradi \& Schwarz \cite{Co93}), HM~Sge, V1016~Cyg and
Hen~2--147 (Solf \cite{So83,So84}; Munari and Patat \cite{Mu93}; Corradi et
al. \cite{Co99}), as well as M2-9 (Smith et al. \cite{Sm05}) and Mz~3
(Guerrero et al. \cite{Gu04}, Santander-Garc\'\i a et al. \cite{Sa04}). 
The latter are sometimes classified as planetary nebulae, but possibly host a
symbiotic nucleus (Schmeja \& Kimeswenger \cite{Sc01}).

In general, the distances to these systems are not well known. About
19 known and possible symbiotic stars have been measured by the {\em
Hipparcos} astrometric mission, and for only four of them the formal
error of their parallax is less than 50\% (Munari \cite{Mu03}). This
uncertainty plagues the estimates of the total number of symbiotic
stars in the Galaxy (Munari and Renzini \cite{MR92}), which scale with
the adopted local density and partnership to galactic stellar
populations (Whitelock and Munari \cite{WM92}). The uncertainty on
distances also precludes the calculation of a number of physical
properties, including the kinematical ages of the nebulae which in
turn can provide the timescales of the mass-loss events which
characterized their recent history (e.g. Corradi et
al. \cite{Co01}). The advent of high spatial resolution instruments
like the HST allows us to tackle the distance problem by attempting to
measure the expansion parallax of the circumstellar resolved nebulae.
This is done by measuring the growth of the nebula in the plane of the
sky, via multi-epoch imaging, and then determining the distance once
their 3-D geometry and kinematics are established using Doppler shift
measurements. To date, such a method has not been used on any
symbiotic nebula, although it has been applied to a number of
relatively close planetary nebulae (NGC~6543, Reed et al. \cite{Re99};
IC~2448, NGC~6578, NGC~6884, NGC~3242 and NGC~6891, Palen et
al. \cite{Pa00}). Note, however, some recent criticisms of the
expansion parallax method (Mellema \cite{Me04}, Sch\"onberner et
al. \cite{Sc05}).

Henize~2-147 (Hen~2--147, also named as PN G327.9-04.3 or V347 Nor) is
a symbiotic Mira with a pulsation period of $\sim$375 days
(Whitelock \cite{Wh88}).  Its extended nebula was discovered by Munari
\& Patat (\cite{Mu93}), and modeled as an expanding ring inclined on
the plane of the sky by Corradi et al. (\cite{Co99}). In this paper,
we present a high-resolution spatio-kinematical study of the ring
nebula of Hen~2--147.  HST images taken in 2001 and 2004 reveal its
expansion in the plane of the sky. Integral-field high-resolution
spectroscopy at the VLT provides us with Doppler shifts for all of the
nebular structures, and together with the HST data enables us to
discuss the dynamics of the nebula and to compute an expansion
parallax.  This is compared with an independent distance determined
from near-IR photometry at SAAO using the Mira period-luminosity
relationship. This combination of information provides insights into
the dynamical properties of the nebula, as well as testing the
validity of the derived expansion parallax for this
object. Preliminary results were presented by Santander-Garc\'\i a et
al. (\cite{Sa05}).

\section{Observations and Data reduction}

\subsection{Imaging}

Two sets of [N{\sc ii}] 658.3 nm images of Hen~2--147 were obtained using
the Wide Field Planetary Camera 2 (WFPC2) on the Hubble Space Telescope
(HST) on September 29, 2001 (programme GO9082) and September 25, 2004
(programme GO9965) respectively. The images were obtained through the F658N
filter (central wavelength/bandpass = 659.0/2.2 nm), which transmits the
[N{\sc ii}] line with 78\% efficiency and the H$\alpha$ line with about 2\%
efficiency at the heliocentric systemic velocity of Hen~2--147 (Corradi et
al. \cite{Co99}).  The entire nebula falls inside the PC CCD, which has a
spatial scale of $0''.045$ pixel$^{-1}$.

The HST images (one dataset per epoch) were obtained using a 4-point
dithering pattern in order to properly sample the HST point spread
function (PSF). They were then reduced using the ``multidrizzle
package'' (Koekemoer et al. \cite{Ko02}), an automated dithered image
combination and cleaning task in PyRAF (Python IRAF).  This routine
first creates a bad-pixel mask using calibration reference files, and
performs the sky subtraction. Next, the software arranges the four
WFPC2 CCDs into a mosaic, where geometric distortion effects are
corrected, including time dependent distortions due to CCD rotations
and flexure. Then it removes cosmic rays and creates, for each epoch,
a combined and drizzled image with a sampling scale of 0$''$.045
pix$^{-1}$.  Finally, the two epoch images are aligned with respect to
the centre of expansion of the nebula around Hen~2--147. Charge
transfer efficiency, which decreases with time, is the only effect
which slightly degrades the generally good ($\sim$0.25 pix) overall
astrometric accuracy between images.

\subsection{Integral Field Spectroscopy}

Integral Field Spectroscopy was obtained at Kueyen, Telescope Unit 2 of the
8m VLT at ESO's observatory on Paranal, on April 2, 2004. The entire nebula
was covered by the ARGUS IFU at its highest spatial resolution (314 fibres
each with a size of $0''.3$ projected onto the sky, giving a total of
6$''$.6 by 4$''$.2 field of view).  Two 240~s and one 50~s exposures were
taken with the H651.5 grating, covering the region between 630 and 670 nm,
and then co-added. The resolving power was R=25000 corresponding to a
velocity resolution of 12.3 km~s$^{-1}$ around the H$\alpha$\ and [N{\sc
ii}]\ $\lambda\lambda$654.8,658.3 emission lines. Seeing was 0$''$.65.
The VLT data were kindly reduced by Reinhard Hanuschik through the ESO
pipeline, and then further analyzed using standard IRAF routines.

\subsection{Near-infrared photometry}

Hen~2--147 has been monitored at near-infrared wavelengths since 1981 from
the South African Astronomical Observatory (SAAO) at Sutherland. The
complete list of $JHKL$ observations is presented in Table 2. All but one
were made on the 0.75m telescope with the MkII Infrared Photometer; the
other one (marked with an asterisk in Table 2) was made with the MkIII
Infrared Photometer on the 1.9m telescope. The photometry is on the SAAO
system, as defined by Carter (1990), and is good to better than $\pm0.03$ at
$JHK$ and better than $\pm0.05$ at $L$.  There are 119 observations made
over 24 years, one of which was published by Munari et al. (1992). The first
five years worth of data were discussed by Whitelock (1987).

\begin{figure*}
\center
\resizebox{11.0cm}{!}{\includegraphics{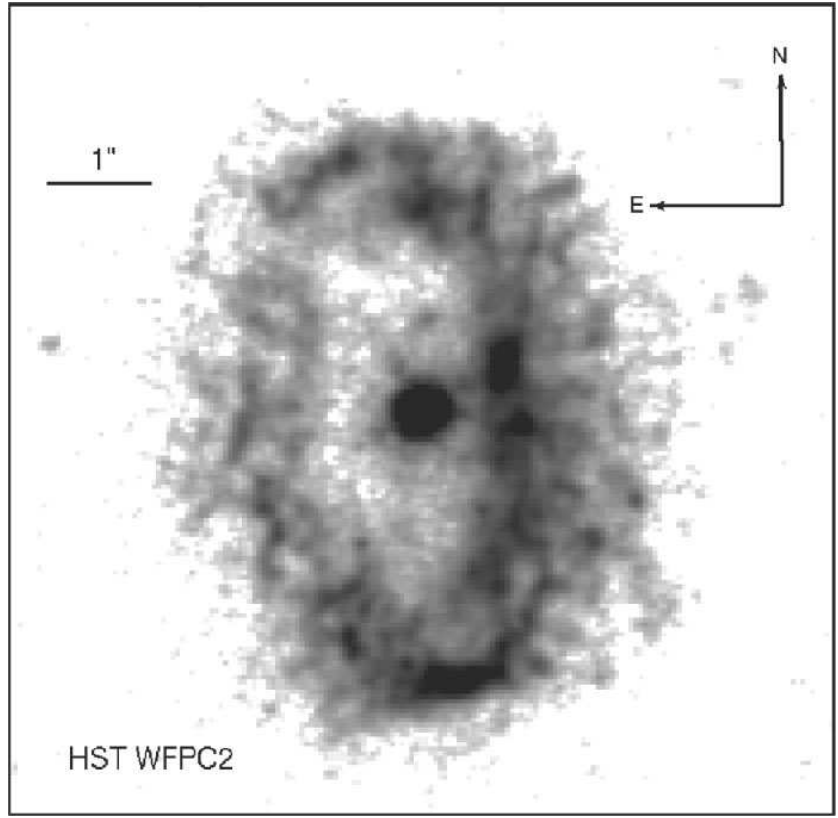}}
\caption{The HST [N{\sc ii}] image of the symbiotic nebula around Hen~2--147
obtained in 2001.}
\label{F1a}
\end{figure*}

\begin{figure*}
\center
\resizebox{11.0cm}{!}{\includegraphics{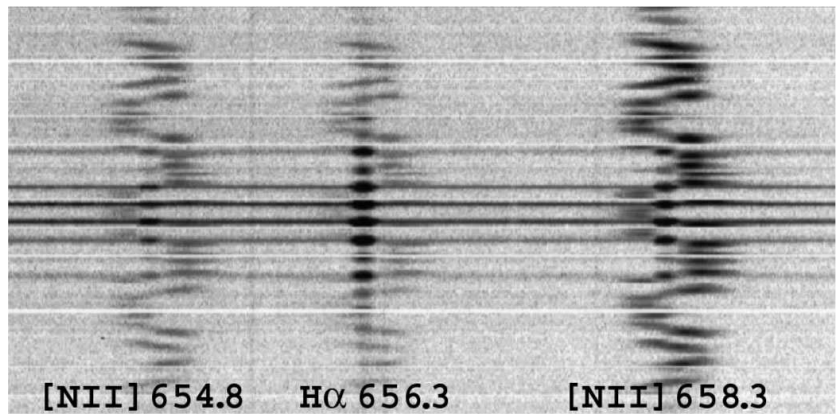}}
\caption{The VLT+ARGUS spectra of Hen~2--147.
Each row is the spectrum of one of the 314 fibres of the ARGUS
Integral Field Unit.  The continuum emission from the core of Hen~2--147
is visible in the middle rows.}
\label{F1b}
\end{figure*}

\section{Morphology and spectra}

The 2001 image of Hen~2--147 is shown in Fig.~\ref{F1a}. The overall
morphology consists of an ellipse of emitting gas detached from an
unresolved central source, confirming the conclusion of Corradi et al.
(\cite{Co99}) from ground based images.  The high spatial resolution of HST
enables us to see that the ellipse is fragmented into a number of clumpy and
filamentary structures; these will be referred to as ``knots'' hereafter.
The overall size of the ellipse is 5$''$.6 by 3$''$.6 measured at 10\% of
the peak intensity of the nebular emission.

The surface brightness is not uniform, as the western and southern knots are
significantly brighter than the eastern ones. Note also that the nucleus is
slightly offset from the centre of the ellipse towards the western knots
(see Fig.~\ref{F66}).

The good seeing and spatial sampling of the VLT spectra, presented in
Fig.~\ref{F1b}, allow us to separate the emission from the central source
from that of the knots.  The central source shows a faint continuum and
relatively narrow H$\alpha$ (brighter) and [N{\sc ii}] 658.3 (fainter)
emission.  In contrast, in the knots [N{\sc ii}] 658.3 is stronger than
H$\alpha$, and all lines are significantly broadened.

\section{Dynamical analysis}

A spatio-kinematical model accounting for the observed morphology,
motions in the plane of the sky, and Doppler shifts of the emission
lines, is presented below.

\subsection{On the plane of the sky}

\begin{figure*}
\center
%WARNING: gif animation, for electronic version only!
\caption{Gif animation of the expansion of the nebula of Hen~2--147 between 
2001 and 2004. Only available in the electronic version of the paper.}
\label{GIF}
\end{figure*}

\begin{figure*}
\center
\resizebox{7.3cm}{!}{\includegraphics{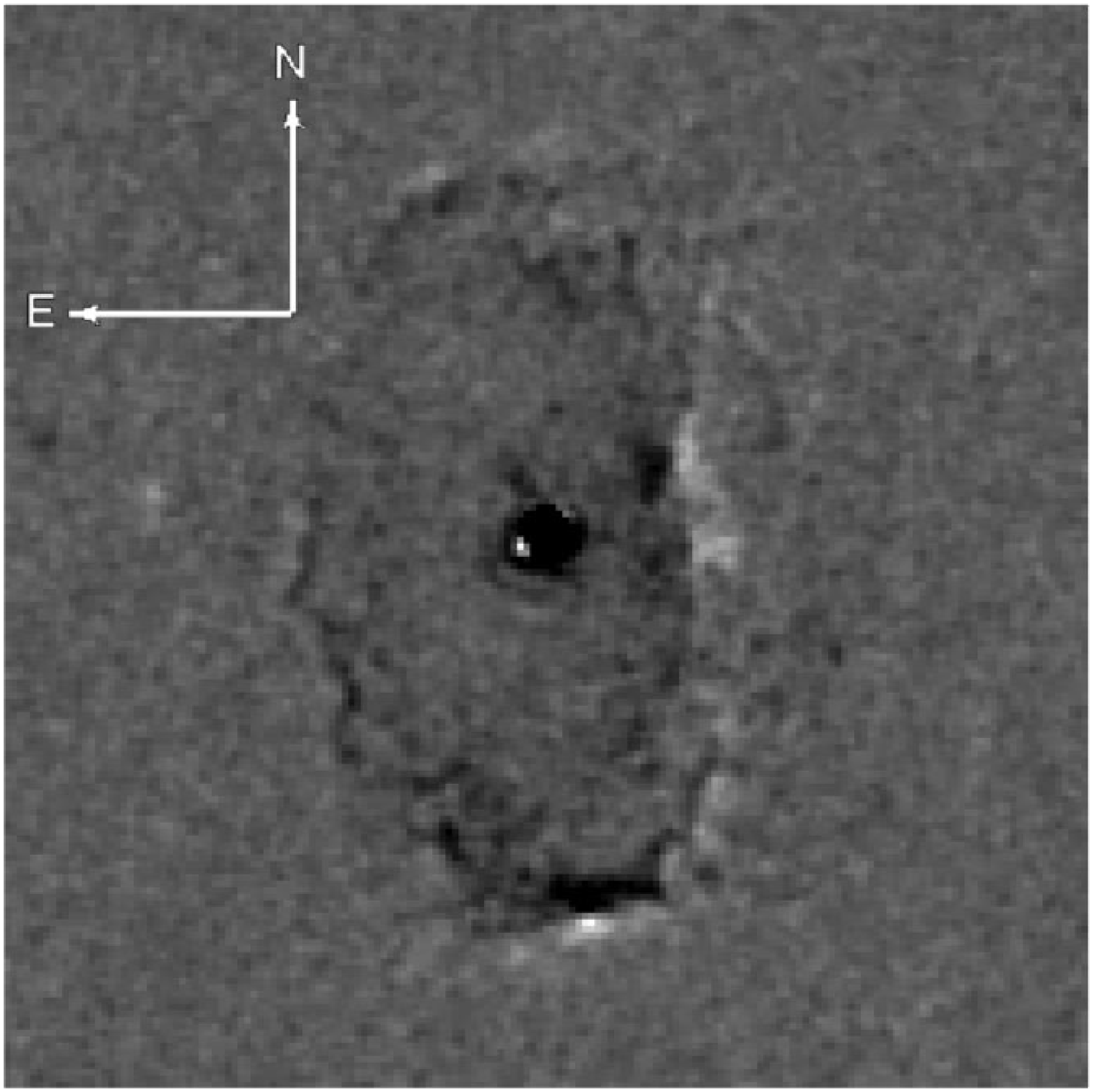}}
\caption{HST [N{\sc ii}]\ WFPC2 residual image, obtained  subtracting the 2001
image from the 2004 one.}
\label{F2a}
\end{figure*}

A simple blinking of the images obtained in 2001 and 2004 (Fig.~\ref{GIF}, a
gif animation available in the electronic version of the paper) reveals the
expansion of the nebula. Fig.~\ref{F2a} shows the residual map resulting
from the subtraction of the 2001 image from the 2004 one, after first
aligning the two images with respect to the expansion centre of the nebula
as defined below. At each radial direction from the centre, positive
residuals (white in the figure) are located outwards while negative
residuals (black) are located inwards.  This indicates that the nebula has
expanded significantly between the two epochs.

In order to quantify the amount of this expansion it is important to define
the expansion centre, as the unresolved core seems not to be located at the
centre of the elliptical nebula. In the hypothesis of purely radial
expansion, the expansion centre can in principle be determined as the
intersection of all the vectors describing the expansion of each knot.
However, an attempt to determine the knots' expansion vectors by measuring
the displacement of their flux-weighted barycentre, or by cross-correlating
their positions between the two epochs, yields uncertain results. Probably
because the knots are diffuse, not well separated from each other, and
they apparently change slightly in shape between the two epochs.

Therefore, the amount of expansion along a number of radial directions was
determined by 1-D cross-correlation of the surface brightness radial
profiles between the two epochs. In order to do this, we have considered
three different possible expansion centres: i) the centre of the ellipse
fitted to the nebula by eye, ii) the central unresolved emission-line core,
and iii) the centre of the ellipse fitted to the nebula by an
emission-weighted least-squares method. For each of these potential centres,
radial surface brightness profiles were extracted at 60 different position
angles (with 6$\degr$ increment steps) in the 2001 and 2004 images, and
cross-correlated to compute the proper motion for each direction.

\begin{figure*}
\center
\resizebox{14.0cm}{!}{\includegraphics{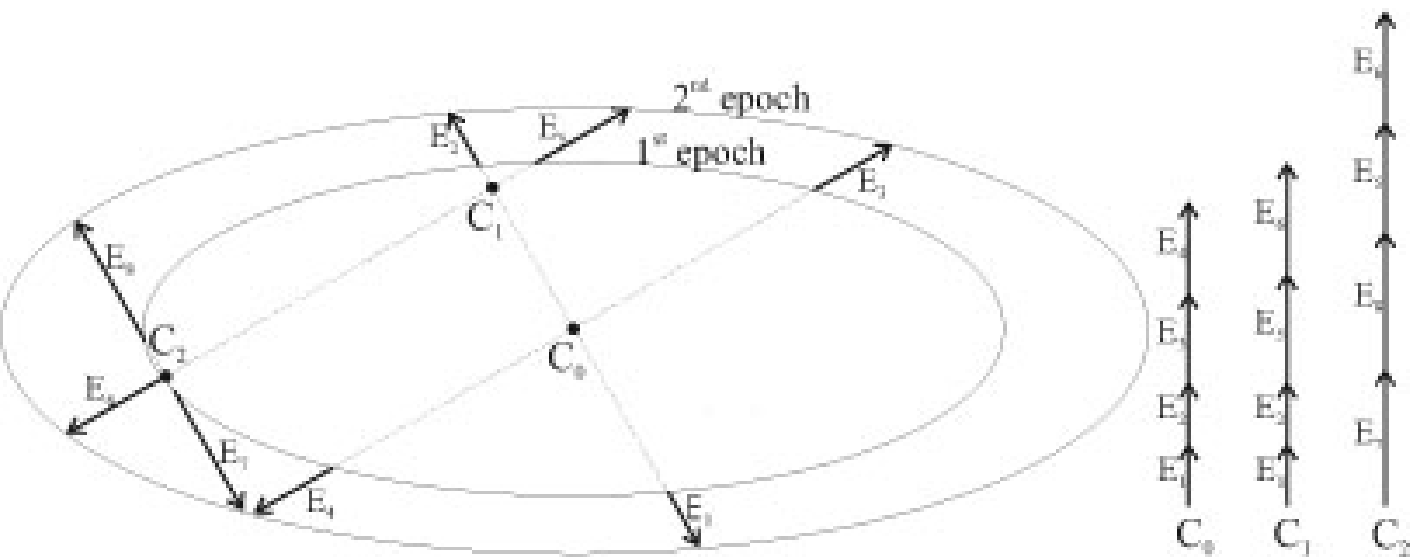}}
\caption{Given an ellipse which expands radially from epoch 1 to epoch 2, the
point which corresponds to the minimum sum of 1-D radial expansions
along several uniformly distributed directions (see the sums for
$C_0$, $C_1$ and $C_2$ in the right side) is the centre of the
ellipse, $C_0$ (i.e. the point which {\it sees} the minimum total
radial expansion).}
\label{F67}
\end{figure*}

\begin{figure*}
\center
\resizebox{7.3cm}{!}{\includegraphics{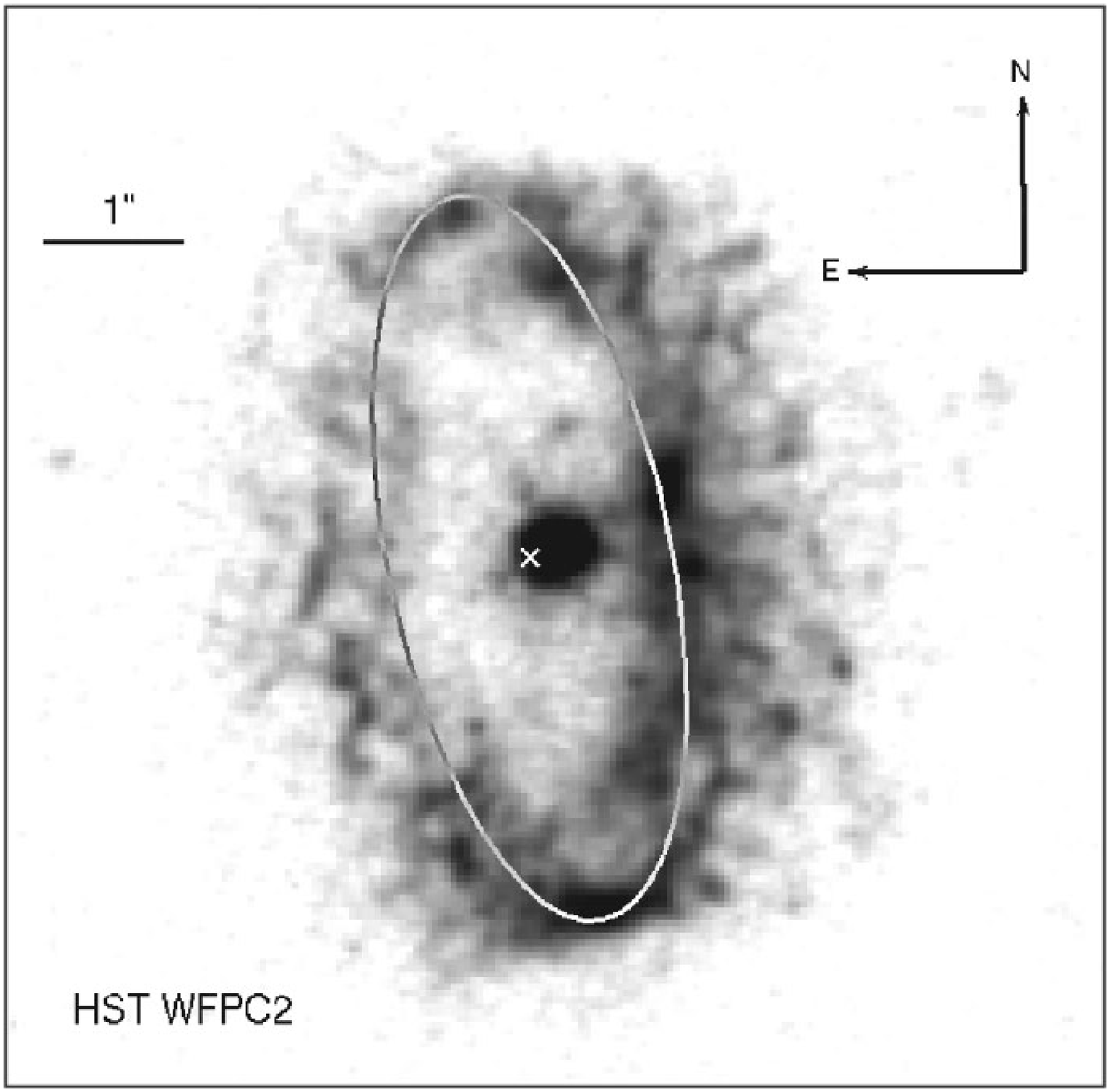}}
\caption{ HST [N{\sc ii}]\ WFPC2 2001 image, with the emission-weighed
least-squares-fitted ellipse and its centre overlaid. The centre is offset
0$''$.2 at P.A.=111$^\circ$ from the nucleus.}
\label{F66}
\end{figure*}

As illustrated in Fig.~\ref{F67}, the centre of a radially expanding
ellipse can be defined as the point which minimizes the sum of the
radial expansions of a number of individual features uniformly
distributed along the ellipse. In the case of Hen~2--147, the sum is
minimum (by a significant 4\%) for the centre of the
least-squares-fitted ellipse, which in the following is therefore
assumed to be the centre of expansion of the nebula. This is located
0$''$.2 $\pm$ 0$''$.05 from the central star toward
P.A.=111$^\circ$, as indicated in Fig.~\ref{F66}. The orientation of
the long axis of this ellipse is P.A. 11$^\circ$.5 $\pm$0$^\circ$.5,
and its size is 5$''$.2 by 2$''$.0 ($\pm$ 0$''$.1). Assuming that the
intrinsic morphology is that of a circular ring, its polar axis would
be inclined 68$^\circ$.6 $\pm$ 0$^\circ$.5 to the line of sight.

Once the expansion centre was determined, the growth of the nebula was
derived using the so-called ``magnification method'' (see Reed et al.
\cite{Re99}). This consists of finding the magnification factor, $M$,
to be applied to the 2001 image, which minimizes the {\it rms} of the
map obtained by subtracting the magnified image from the 2004 image.
Provided that there are no changes in the brightness of each feature
and that the motion is radial, a given feature will therefore
disappear in the subtracted image.

\begin{figure}
\center
\resizebox{7.3cm}{!}{\includegraphics{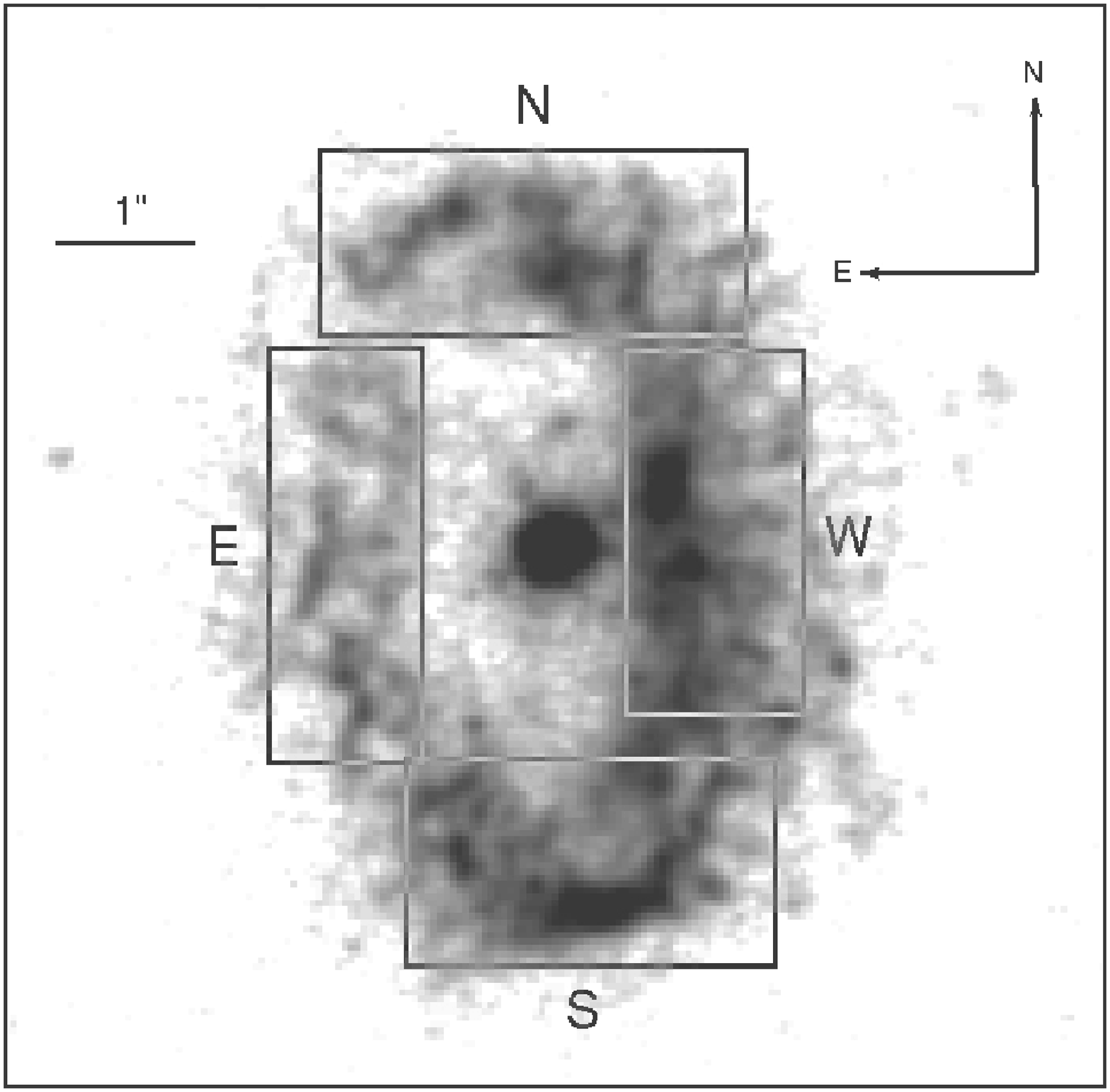}}
\caption{ HST [N{\sc ii}]\ WFPC2 2001 image. The
four regions in which the {\it rms} of the magnification method residual
image were measured are indicated by the black boxes.}
\label{F2b}
\end{figure}

\begin{figure*}
\center
\resizebox{12cm}{!}{\includegraphics{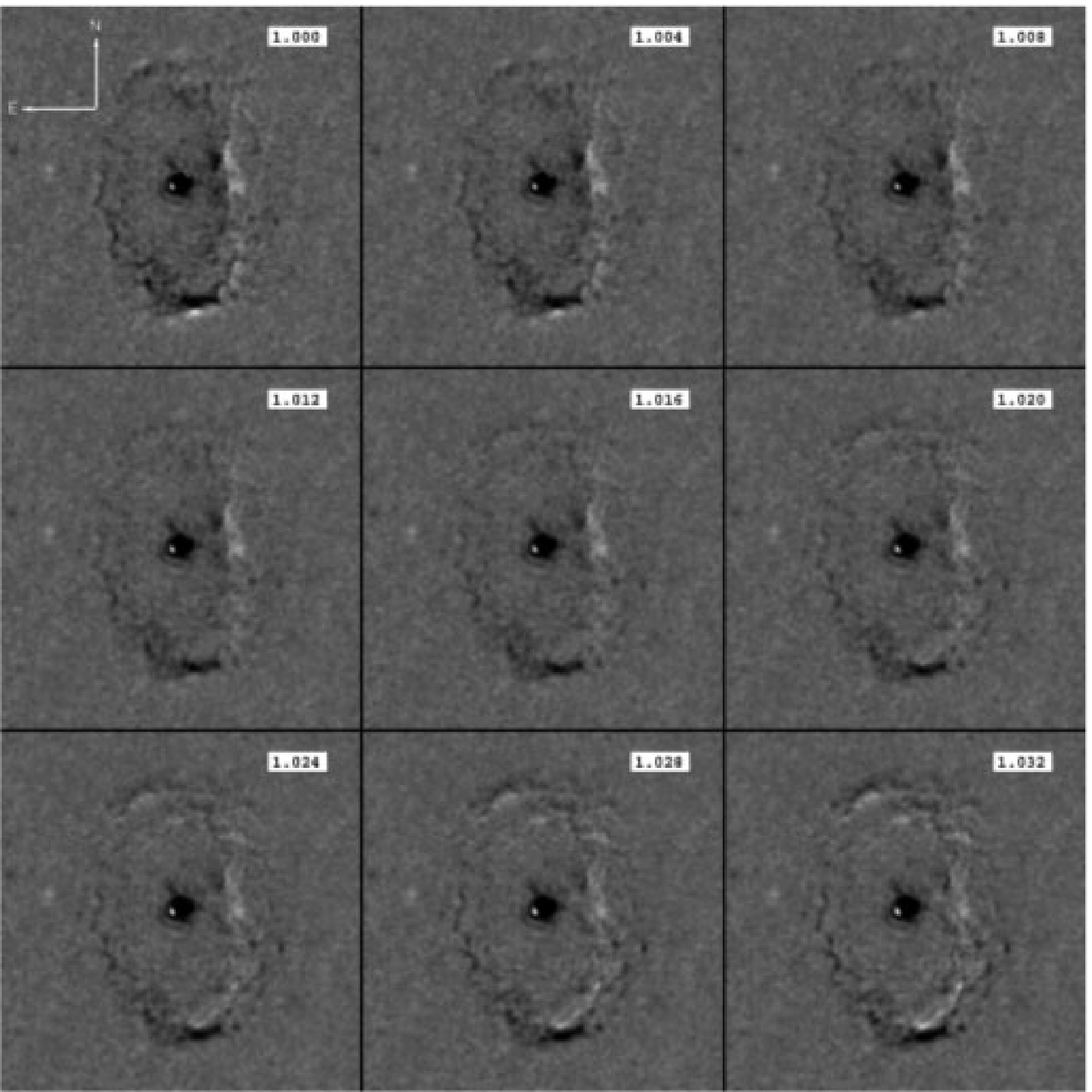}}
\caption{Sample of HST [N{\sc ii}] residual images obtained with
different magnification factors (see labels). The residuals for the north,
south and east regions are minimized with magnification factors of 1.013,
1.015 and 1.017 respectively, while the western region has grown much
faster, by a factor 1.029.  Because the nucleus is offset from the centre of
expansion of the nebula (see text) the white point marking the centre of the
ellipse is displaced from the central stellar image.}
\label{F3}
\end{figure*}

The complex structure of the nebula and the evolution of the shape and
brightness of knots prevents the analysis of each individual
feature. For this reason, we have considered the regions of the nebula
along the four cardinal directions as a whole (see
Fig.~\ref{F2b}). Magnification factors $M$ spanning the interval from
1.000 to 1.050 were applied. The $M$ factors giving the minimum {\it
rms} in the subtracted image are 1.013, 1.015, 1.017 and 1.029, for
the north, south, east and west region, respectively; with an
uncertainty of $\pm$0.002 in each case. An illustrative sample of
residual images for different $M$ factors is shown in Fig. ~\ref{F3}.

Thus, the growth of the nebula is uniform with an average value of
$M=1.015\pm$0.002 except in the western direction, corresponding to the
brightest knots, where it has grown by about twice this amount. Note also
that for any $M$ the residuals in the southern region are negative, due to a
decrease of $\sim$10\% in the surface brightness of the knots from 2001 to
2004. The rest of the nebula shows no significant changes in surface
brightness between the two epochs.

\subsection{Along the line of sight}

\begin{figure}
\center
\resizebox{7.3cm}{!}{\includegraphics{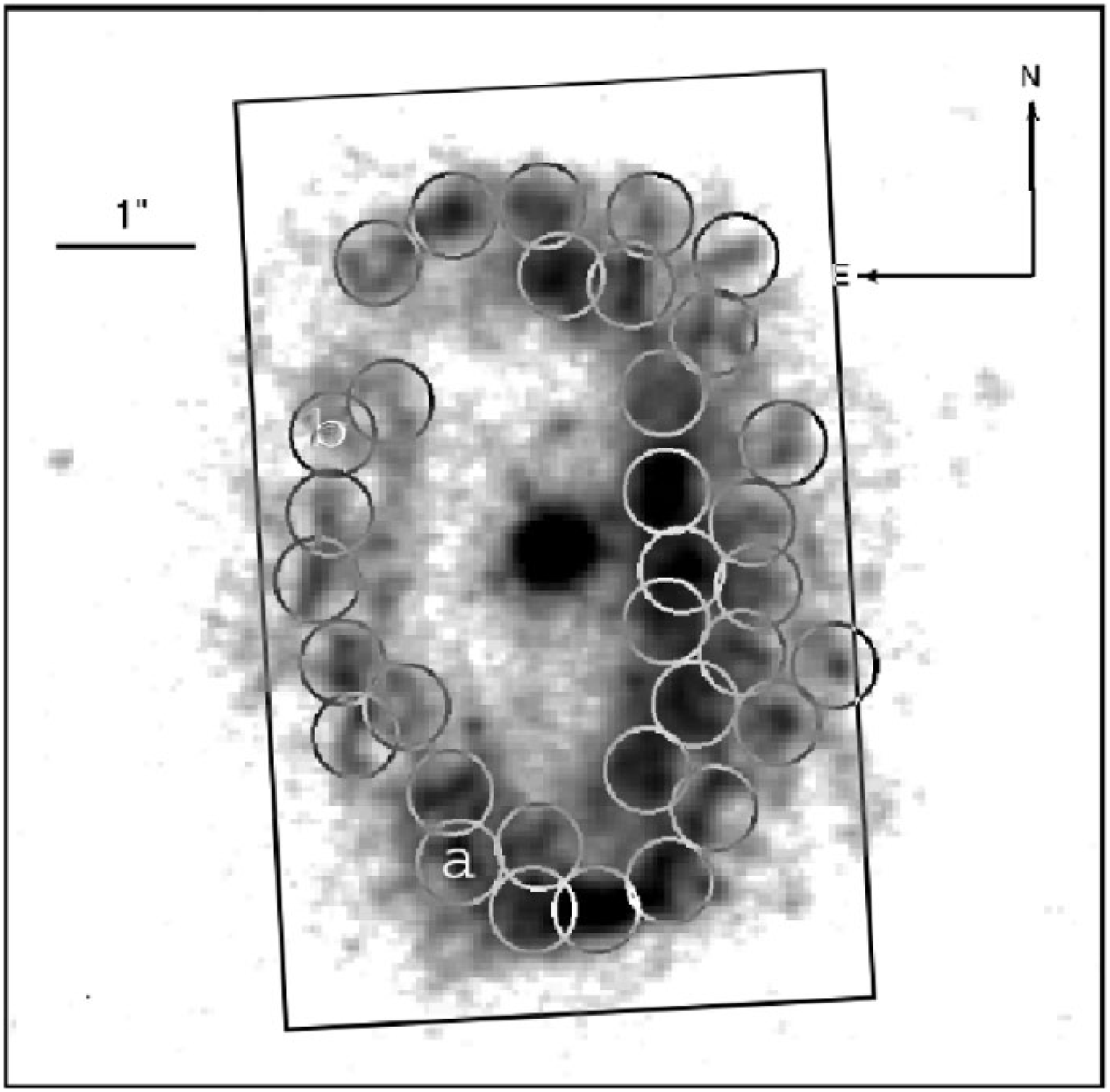}}
\caption{The ARGUS field of view, indicated by the black rectangle 
superimposed on the HST image. The 34 knots whose spectra were analysed are
overlaid as 0$''$.6 diameter circles. The H$\alpha$ and [N{\sc ii}] 658.3
line profiles of the core and of sample knots labeled as {\it a} and {\it b}
are plotted in Fig.~\ref{F9}.}
\label{F4}
\end{figure}

The HST image was matched to the IFU array of the VLT+FLAMES, and the
spectra corresponding to 34 knots and the core of Hen~2--147 were extracted
(Fig.~\ref{F4}). Depending on the position and size of each knot either the
spectrum from a single fibre (0$''$.3) was taken or the spectra of up to 4
fibres were summed to give sufficient signal to noise.

\begin{figure*}
\center
\resizebox{12cm}{!}{\includegraphics{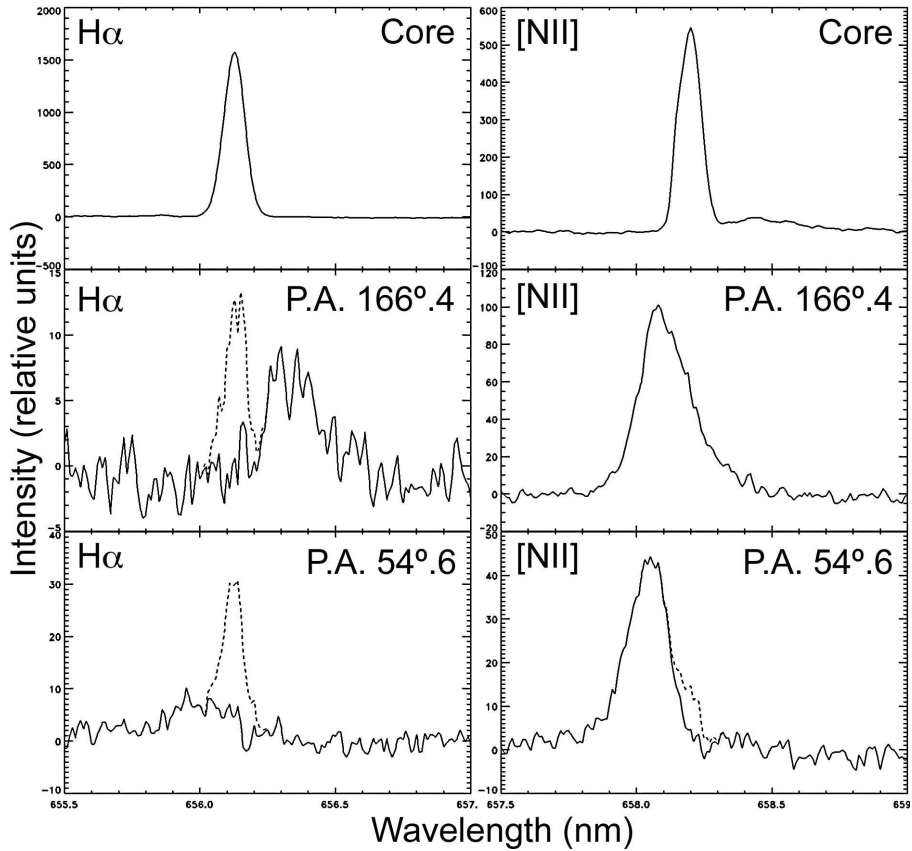}}
\caption{{\bf Top}: H$\alpha$ and [N{\sc ii}] 658.3 profiles of the unresolved core of 
Hen~2--147.
{\bf Middle}: H$\alpha$ and [N{\sc ii}] 658.3 profiles of the knot labelled
as {\it a} in Fig.~\ref{F4}, before (dashed line) and after (solid line) 
subtraction of the contamination from the core emission. Note the asymmetrical 
line profile. {\bf Bottom}: Profiles of the knot {\it b} in Fig.~\ref{F4}.}
\label{F9}
\end{figure*}

Both the H$\alpha$ and [N{\sc ii}]658.3 line profiles of the core and some
illustrative knots are presented in Fig.~\ref{F9}.

In the knots close to the centre as well as in the faintest ones (owing to
the small size of the nebula) the contamination from the core emission is
non negligible. This contamination was removed by scaling the core spectrum
and subtracting it from the spectra of the knots, so as to cancel the
H$\alpha$ component from the core, which is always clearly distinguishable
given its specific redshift, line width, and the fact that H$\alpha$ is much
stronger than [N{\sc ii}] in the core (while the opposite is true for the
nebula). An example of the removal of the core emission is also illustrated
in Fig.~\ref{F9}.

\begin{figure*}
\center
\resizebox{12cm}{!}{\includegraphics{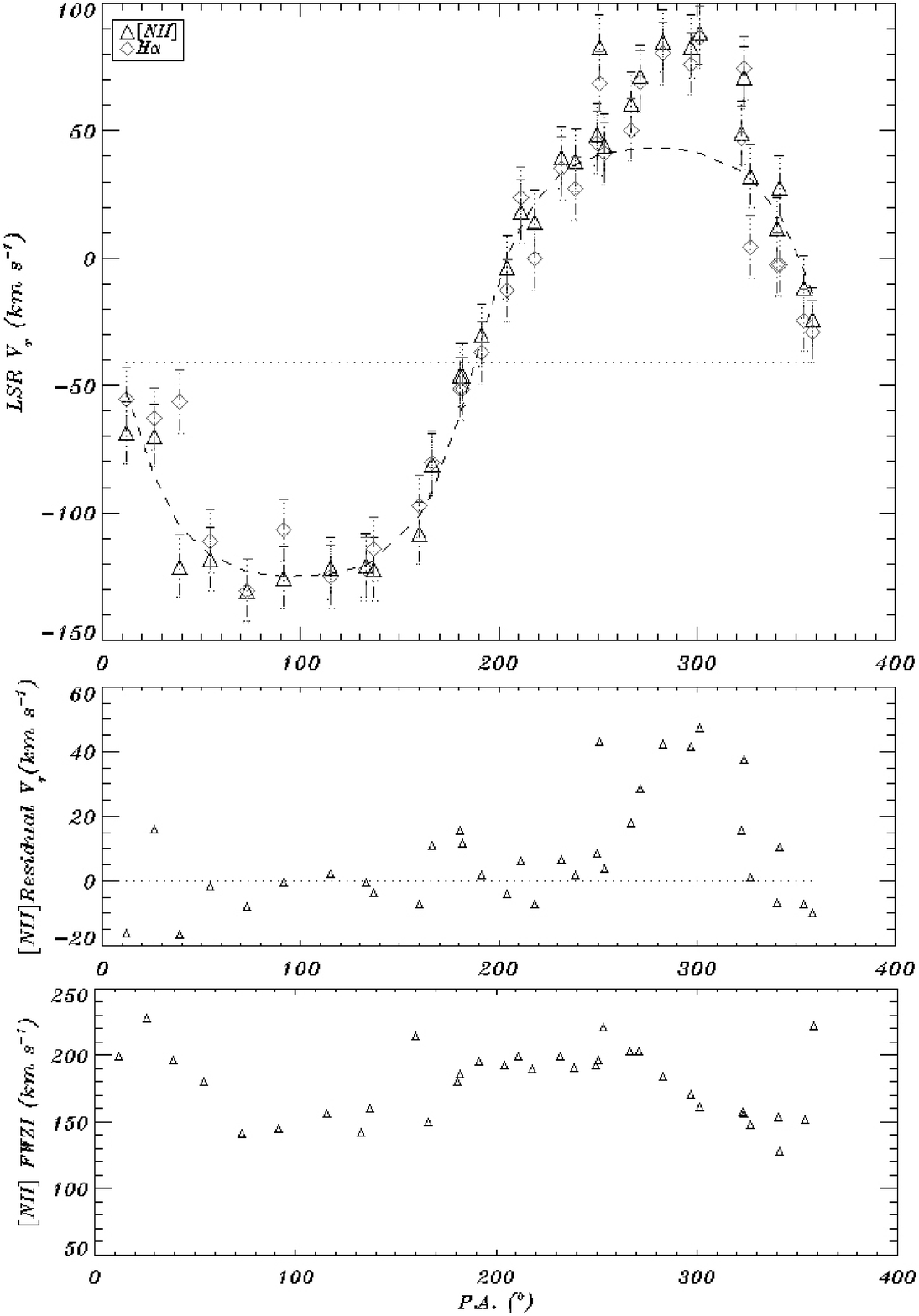}}
\caption{{\bf Top}: LSR radial velocity of the barycentre of both
H$\alpha$ (gray diamonds) and [N{\sc ii}]\ (black triangles) emission
lines of the 34 studied knots of the nebula of Hen~2--147. Error bars
indicate the formal error in the determination of the velocity
centroids. The dashed line represents the expanding ring model (see
text), while the dotted line corresponds to the LSR radial velocity of
its centre. {\bf Middle}: Residual radial velocity defined as the
excess of the [N{\sc ii}]\ data V$_r$ with respect to the ring model.
{\bf Bottom}: Full Width at Zero Intensity of the [N{\sc ii}]\ data.}
\label{F6}
\end{figure*}

Doppler shifts were derived for the barycentre of each line profile in every
knot, as well as for the core. The LSR radial velocity of the central source
is v$_{LSR} = -47 \pm3$ km s$^{-1}$.  In the nebula, the hydrogen and
nitrogen ions are found to expand at the same pace within the observational
uncertainties.  The radial velocity pattern as a function of the position
angle is presented in Fig.~\ref{F6}. This is modelled assuming an expanding
circular ring of gas inclined at $68\pm2\degr$ to the line of sight,
corresponding to the fit of the ellipse in the image (see section 4.1). The
best-fitting de-projected expansion velocity of the ring is 90~km~s$^{-1}$,
with centre at $V_{LSR} = -41$ km~s$^{-1}$ (and thus slightly offset from
the systemic velocity of the nebular core). The residuals of the fitting are
shown in the middle panel of Fig.~\ref{F6}.  The points between
P.A.=220$\degr$ and 330$\degr$ were excluded from the fitting, given their
peculiar velocity pattern in the plane of the sky and considering that if
they are included the quality of the fit decreases.  Note that their excess
radial velocity corresponds to an excess in their velocities in the plane of
the sky; we will come back to this point in Sect.~5.2.

\begin{figure}
\center
\resizebox{7.3cm}{!}{\includegraphics{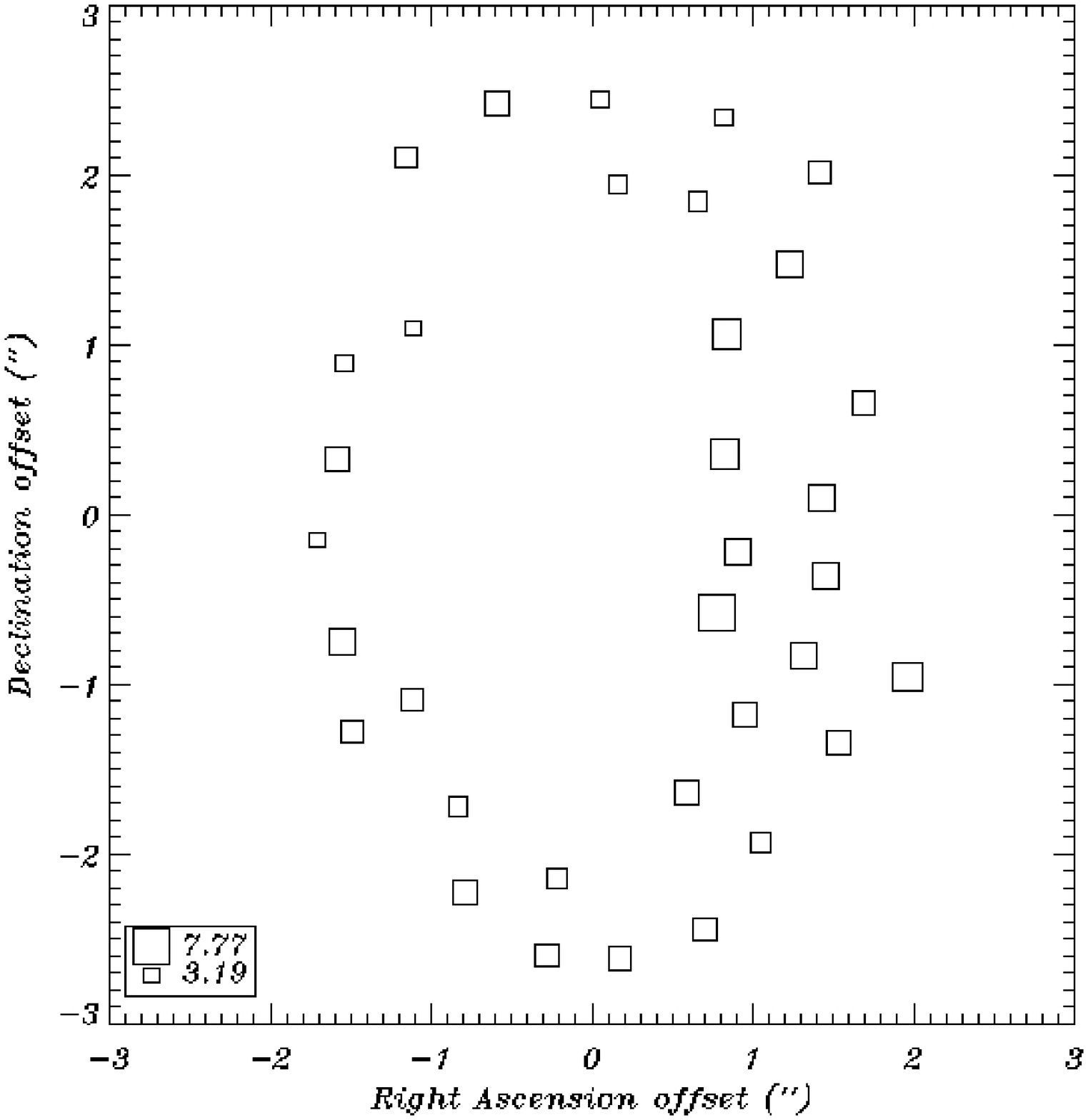}}
\caption{[N{\sc ii}]/H$\alpha$ ratio in the 34 studied knots over the
nebula; the sizes of the squares are proportional to this ratio.}
\label{F8}
\end{figure}

A remarkable property of the line profiles of the knots is their large width,
which is discussed in detail in next section. Furthermore, the [N{\sc ii}] to
H$\alpha$ flux ratio is always very large, especially in the western region
(Fig.~\ref{F8}). Finally, most of the [N{\sc ii}] profiles are slightly
asymmetrical, presenting redward (more common) or blueward wings.

\subsection{Profile broadening}

Both the [N{\sc ii}] and the H$\alpha$ lines in all knots have a full
width at zero intensity (FWZI) between 150 and 200~km~s$^{-1}$ (with a
mean value of $\sim160$~km~s$^{-1}$, see Fig.~\ref{F6}, bottom
panel). Which is much larger than the instrumental or thermal
broadening, as previously noted by Corradi et al.  (\cite{Co99}).  In
this section we discuss several hypotheses that might account for this
property.

The combination of the observed morphology and velocity field
demonstrates that the nebula of Hen~2--147 is a ``2-D" ring inclined
on the plane of the sky. This excludes the possibility that the wide
range of velocities forming the line profiles at each knot position
come from regions of the nebula projected along the same line of
sight, each with relatively small intrinsic velocity dispersion, but
projecting at different angles (as, e.g., in the case of a projected
ellipsoid). We can also exclude the possibility that the broadening is
related to intrinsic gas emission mixed with dust scattering, which
would result in additional components to the Doppler shifts.

Therefore the velocity dispersion is very likely to be intrinsic to each
knot.  One possibility is that during the nebular evolution the knots have
expanded into a surrounding medium of lower density, temperature and
pressure.  The fast relative flow of the ambient medium along the sides of
the knots would reduce the pressure, Rayleigh-Taylor instabilities would
break the knots into smaller filaments and blobs, and Kelvin-Helmholtz
instabilities mix the knot material with the ambient flow (Soker \& Regev
\cite{So98b}).

Given the explosive nature of symbiotic novae like Hen~2--147, and the high
velocities observed, another natural explanation for the large line widths
would be the presence of shocks. They might originate as the result of
interactions between the slow Mira wind and the fast ejecta produced in the
outburst. The large [N{\sc ii}] to H$\alpha$ ratio throughout the nebula
(Fig.~\ref{F8}) supports this possibility.  According to Hartigan et
al. (\cite{Ha87}), the velocity of a radiating bow shock is independent of
its density, shape, ionisation state, abundances and temperature, and equal
to the full width at zero intensity of the line profile.  Although the knot
geometry in the case of Hen~2--147 is far more complex and varied than that
of the model analysed by Hartigan et al.  (\cite{Ha87}), the predicted
effects are qualitatively similar: the line profiles widen and present
asymmetric shapes throughout the nebula. Several nebulae in the literature
present similar broad and irregular [N{\sc ii}] and [S{\sc ii}] profiles
which can be explained by shock excitation, as in the case of P Cygni
(Barlow et al. \cite{Ba94}).

\section{The distance}

\subsection{Period-Luminosity relationship for the Mira of Hen~2--147}

\begin{figure}
\center
\resizebox{8.0cm}{!}{\includegraphics{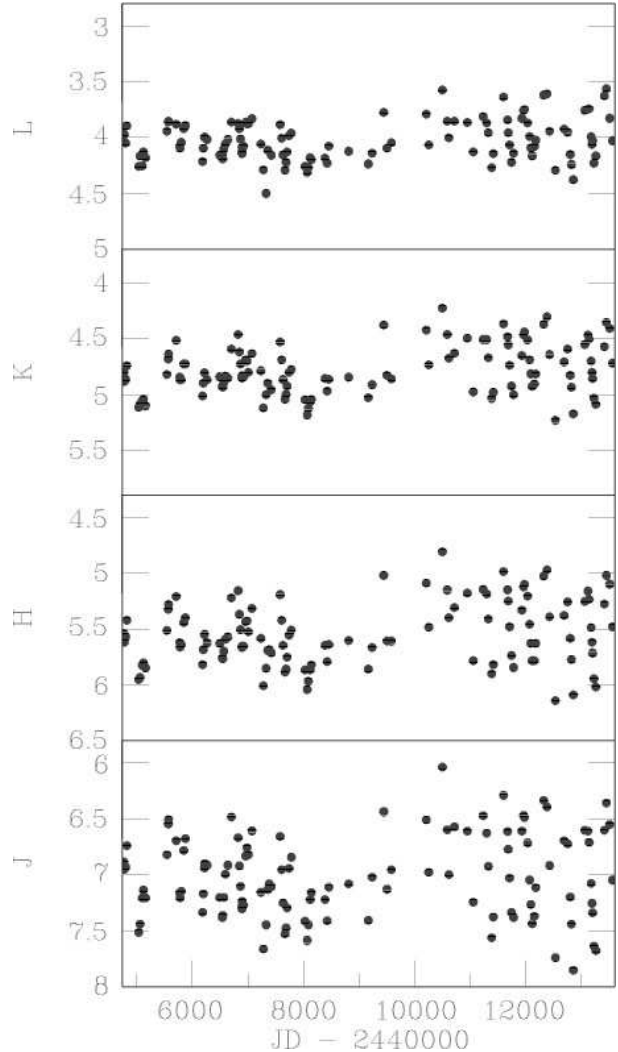}}
\caption{Near-infrared light-curves for Hen~2--147.}
\label{fig_jhkl}
\end{figure}

\begin{figure}
\center
\resizebox{8.0cm}{!}{\includegraphics{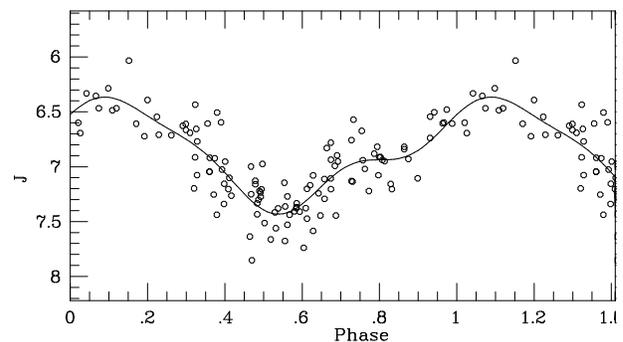}}
\caption{$J$ light-curve for Hen~2--147, phased on the 373 day period with 
an arbitrary zero point; the best fitting 3rd order sine-curve is also
shown.}
\label{fig_j}
\end{figure}

The $JHKL$ light-curves of Hen~2--147 spanning the last 25 years are
presented in Fig.~\ref{fig_jhkl}.  The mean light levels vary by a few
tenths of a magnitude, as they typically do among non-symbiotic O-rich Miras
(e.g.  Whitelock et al. \cite{Wh00}). Note, however, that these light-curves
show no evidence of obscuration events of the type very commonly observed in
light-curves of symbiotic Miras (e.g. Whitelock \cite{Wh87}).  A Fourier analysis of these data indicates a pulsation period of 373 days 
with maximum light at $J$ occurring on JD 2450476; Fig.~\ref{fig_j} shows the $J$ 
light-curve phased at that period. The
Fourier mean magnitudes and peak-to-peak amplitudes are given in Table 1.
These amplitudes are within the range shown by non-symbiotic Miras with
similar pulsation periods (Whitelock et al. \cite{Wh00}).

\begin{table}
\begin{center}
\begin{tabular}{ccccc}
\hline
& $J$ & $H$ & $K$ & $L$ \\
mag & 6.85 & 5.43 & 4.71 & 3.95 \\
$\Delta$ & 0.89 & 0.70 & 0.52 & 0.40\\
\hline
\end{tabular}
\end{center}
\label{means}
\caption{Mean magnitudes and pulsation amplitudes for Hen~2--147.}
\end{table}

If we assume that the Mira in Hen~2--147 has similar characteristics to its
solitary counterparts then it is possible to derive a distance from the Mira
period-luminosity (PL) relation (Feast et al. \cite{Fe89}), after
correcting for extinction. Whitelock et al. (\cite{Wh00}) give period-colour
relations for unreddened oxygen-rich Miras which can be used to estimate the
intrinsic colours of the Hen~2--147 Mira. Then the extinction $A_V$ can be
estimated from each colour, assuming the interstellar reddening law (Glass
\cite{Gl99}) is applicable to both circumstellar and interstellar reddening.

The following values of $A_V$ are derived for the colours mentioned:
$A_V=4.22\ (J-H), \ 3.97\ (H-K),\ 4.89\ (K-L)$ and $4.07\ (J-K)$. The value from
$(K-L)$ is larger than the others and we might reasonably assume that,
in common with other symbiotic Miras, there is a slight excess at $L$. We
therefore assume that the combined interstellar and circumstellar extinction
is $A_V=4.1$ mag. The interstellar extinction can be estimated following the
procedure outlined by Drimmel et al. (\cite{Dr03}) to give $A_V=1.4$ mag
(including their rescaling procedure) or $A_V=1.8$ mag (without rescaling).
This indicates that a significant fraction of the observed extinction is
circumstellar. Therefore our assumption that the circumstellar material
obeys the same reddening law as the interstellar material could be
important.

Correcting the observed mean $K$ from Table~\ref{means} for reddening we
have $K_0=4.71-0.37=4.34$. Using the PL from Feast et al. (\cite{Fe89}),
with a distance modulus for the LMC of $(m-M)_0=18.5$ mag, we find
$M_K=-7.94$ for a 373 day variable. The distance is therefore $2.9\pm 0.4$
kpc, where the error is statistical. It is difficult to quantify systematic
errors due to possible differences in symbiotic and non-symbiotic Miras or
between circumstellar and interstellar extinction, but these will increase
the uncertainty.

The strength of the K{\sc i} resonant doublet at 7665, 7699~\AA, offers an
independent, direct measure of the amount of material along the line of
sight toward Hen~2--147. We use the calibration by Munari \&
Zwitter (\cite{Mu97}) relating the reddening to the equivalent width of
7699~\AA\ line (the 7665~\AA\ being usually affected by telluric absorption
lines). We have measured KI lines on a deep FEROS spectrum we obtained with
the 2.2m telescope at ESO (La Silla) on February 2003. The interstellar
lines show two components. The bluer one is located at heliocentric velocity
$-$37($\pm$1)~km/sec, and has an equivalent width of 0.178~\AA\
corresponding to a reddening $E_{B-V}$=0.69. The redder component is located
at heliocentric velocity $-$21($\pm$2)~km/sec, and has an equivalent width
of 0.065~\AA\ corresponding to a reddening $E_{B-V}$=0.23. The total reddening
affecting Hen~2--147 is therefore $E_{B-V}$=0.92 or $A_V=2.85$ mag. With this 
determination of the total extinction between us and the Mira the distance refines to
3.0$\pm$0.4~kpc.

\subsection{Expansion parallax}

The expansion parallax of the nebula of Hen~2--147 can be determined in a
straightforward way from the present data assuming 1) the circular ring
model of Section 4.2, 2) that the knots of Hen~2--147 expanded radially and
suffered no significant acceleration in the period between 2001 and 2004,
and 3) that we are observing matter motions (and not a shock expansion) in
the plane of the sky. Under these circumstances, a given knot $k$ at an
angular distance $\alpha$ in the plane of the sky in the 2001 image will
have moved to $\alpha+\Delta \alpha$ in the 2004 one. The magnification
factor $M$ can be defined as $M =\frac{\alpha+\Delta \alpha}{\alpha}$, so
that the angular expansion in the plane of the sky is $\Delta \alpha =
\alpha (M-1)$.  Let $\theta$ be the angle of a given point of the ring, in
the plane of the sky, measured from the minor axis of the ellipse (i.e.
latitude of the ellipse in the plane of the sky).  Considering the expanding
inclined ring model, a given knot at an angle $\theta$ (measured from the
minor axis of the ellipse, in the plane of the sky) which moves at a spatial
velocity $V$ will therefore show an expansion $\Delta d$ in the
plane of the sky, during an interval of time $\Delta t$, given by 
\begin{equation}
\Delta d = V\ \Delta t \sqrt{\frac{\cos^2{i} \ (1 + \tan^2{\theta})}{1 + \tan^2{\theta} \cos^2{i} }}
\end{equation}
Thus, the
distance $D$ from the observer to Hen~2--147 is simply 
\begin{equation}
D(kpc) = \frac{206265}{3.086 \ 10^{16}}\ \frac{\Delta d(km)}{\Delta \alpha('')}
\end{equation}

We then compute the distance for each of the four regions analysed in
section 4.1, finding values of 1.84$\pm$0.34, 1.60$\pm$0.27, and
1.84$\pm$0.46~kpc for the north, south, and east regions respectively.  As
discussed in Sections~4.1 and 4.2, the western side of the nebula presents
excess velocities both in the plane of the sky and along the line of sight.
Adding these excess velocities to the spatio-kinematical model adopted, a
distance, 1.21$\pm$0.53~kpc, similar to that found for the other regions is
determined.  This result supports the idea that the adopted geometry (and
inclination) applies also to the western portion of the nebula, and that gas
in different regions has not suffered significant or differential
acceleration between 2001 and 2004.  Averaging the determinations above, we
obtain $D=$1.5$\pm$~0.4~kpc for the expansion parallax of Hen~2--147, a
result which does not change if any of the other possible centres of the
nebula is adopted (Sect.~4.1).

\subsubsection{The effects of shocks on the expansion parallax}

The Mira PL relation and the expansion parallax provide apparently
inconsistent results.  One way to reconcile them is to consider that shocks
play a major role in the application of the expansion parallax method.  In
the case of a strong shock, the apparent expansion in the plane of the sky
that we observe in the images would not represent the expansion of matter,
but rather the (faster) expansion of the shock front through the material.
On the other hand, along the line of sight (spectra) we would be measuring
the Doppler shift of the gas motion. If shocks are important, the expansion
parallax method would underestimate the distance (Mellema, \cite{Me04},
Sch\"onberner \cite{Sc05}).

According to the bow shock models of Hartigan et al. (1987), the shock
velocity $V_s$ is given by the FWZI of the nebular emission lines. In
the case of Hen~2--147, that implies $V_s=$160 km s$^{-1}$ (see
Fig.~\ref{F6}, bottom panel). Under this hypothesis, in equation (1)
the gas expansion velocity should be replaced by the shock velocity,
and as a consequence the distance to Hen~2--147 derived from the
expansion would increase to $D=$2.7$\pm$0.5~kpc.

\section{Discussion and conclusions}

Previous studies of Hen~2--147 showed that this symbiotic system
contains an oxygen-rich Mira and an extended ionized nebula, which
appeared to be an inclined expanding ring of gas ejected from the
system a few hundred years ago (Corradi et al. \cite{Co99}).  In this
work, we have gathered high quality multi-wavelength observations
which allow us to investigate in more detail the structure and
dynamics of the nebula.  As often happens in science, this closer look
reveals several puzzling properties which have to be explained as part
of our endeavour to understand the nature of the object and to
determine its basic properties, including its distance via the
expansion parallax of the nebula.

The near-IR photometry has allowed us to study the light
curve of the Mira within Hen~2--147. Its pulsation period is 373
days and it lacks the obscuration events typical of many other D-type
symbiotic stars. The [N{\sc ii}]\ HST images, of unprecedented spatial
resolution, show the nebula surrounding Hen~2--147 as an ellipsoid
which is fragmented into numerous diffuse knots and filaments of
different surface brightness. The multi-epoch imaging clearly resolve
the expansion of the nebula in the plane of the sky.  Doppler-shift
measurements from VLT integral-field spectroscopy, in combination with
the HST images, demonstrate that the intrinsic geometry of the nebula
is indeed that of a circular, knotty ring of ionized gas with a
5$''$.2$\pm$0$.''$2 diameter, inclined 68$\pm$2$\degr$ to the line of
sight, and expanding at $\sim$90 km s$^{-1}$.

Whatever the details of the dynamics of the nebula are, it is clear
that the circumstellar gas distribution within Hen~2--147 is highly
aspherical. It seems possible that the plane within which the nebula
is concentrated actually coincides with the orbital plane of the
system, although there is no direct evidence for this. Some models
predict gravitational focusing of the Mira mass-loss toward the
orbital plane (Mastrodemos \& Morris \cite{Ma99}; Gawryszczak et
al. \cite{Ga03}), while others suggest that equatorially compressed
AGB winds can form if rotation of the star is near its critical
break-up rate (Garc\'\i a-Segura et al. \cite{Ga99}).  In such a case,
any fast wind or ejecta from the outbursting white dwarf companion
would produce strong shocks in the orbital or rotational plane, these
could manifest themselves in the type of nebula observed around
Hen~2--147. It is, however, remarkable that there is no evidence at
all for ionized gas at moderate and high latitudes above/below the
plane of the ring. It thus appears that, even in widely spaced binary
systems with orbital periods of tens or perhaps even hundreds of years
like symbiotic Miras, binary interactions have a strong influence on
the mass-loss geometry of the system. Note that several planetary
nebulae, such as K~1--9 and ScWe~3 (Schwarz et al.  \cite{Sc92}), also
appear as thin rings whose formation mechanisms might be similar to
that of Hen~2--147. The alternative, that the equatorially collimated
outflows are actually ejecta from the white dwarf companion,
interacting with a much less dense (and less collimated) circum-binary
AGB wind, is less likely.

The new and rather puzzling information revealed by the observations
discussed here is described below.

First, the core is offset from the centre of the expansion of the
nebula, both spatially and in terms of its radial velocity.  This
effect has been found in other objects, possibly because all of them
are binary systems. In addition to Hen~2--147, other examples may be
GG Tau, a young multiple-star system (Krist et al. \cite{Kr02}),
MyCn~18, a planetary nebula which is suspected of containing a binary
nucleus (Bryce et al. \cite{Br04}), and several other planetary
nebulae studied by McLean et al. (\cite{Mc03}). One can try to explain
the offset considering momentum conservation and an inhomogeneous
circular nebula.  The nucleus would recoil back after the ejection,
receding from the denser, more massive knots. Even if ionisation is
mainly due to shocks and the western knots are not the more massive
ones, at the distance discussed earlier and with a typical symbiotic
system mass of 2 M$_\odot$, the nucleus would need to recoil back at
$V$ $>$ 10 km s$^{-1}$ after ejecting M $>$ 0.2 M$_\odot$ towards the
east. This is at least one order of magnitude larger than the total
mass lost by the Mira since the nova outburst. Instead, the property
can be satisfactorily accounted for by orbital interaction in an
eccentric binary system (Soker et al. \cite{So98a}). The difference in
the Mira orbital velocities in apastron and periastron could give
rise to an asymmetry in mass loss, averaged over an entire
orbit. This, after the reshaping of the Mira winds by the white dwarf
during an outburst, would eventually lead to a significant offset of
the nucleus from the centre of the nebula. A similar scenario has been
suggested for MyCn~18 by Soker et al. (\cite{So98a}) and leads to a
very speculative orbital period of a few hundreds of years, which is
typical of symbiotic systems containing a Mira.

Secondly, a most notable property of the outflow of Hen~2--147 is the
extreme broadening of the emission lines in the spectra of the knots
all over the nebula (FWZI $\sim$160 km s$^{-1}$). The broadening is
significantly larger than the ``average'' (intensity-weighted)
expansion velocity itself. This has important implications for the
understanding of the dynamics of the outflows, and in the
determination of the distance.  Two possible explanations for the line
broadening are outlined below. ($i$): the expansion of the outflow in
a high density medium which undergoes strong cooling might develop a
thin front layer which is fragmented into clumps of different density.
The flow velocity along the walls of these structures could be of the
same order of the expansion velocities themselves (Icke, private
communication); ($ii$): the presence of a strong shock, traveling
outwards through the nebula, as a result of the interaction between
the slow Mira wind and the fast white dwarf ejecta produced during the
nova explosion. In fact, although the high [N{\sc ii}]/H$\alpha$ flux
ratio does not imply unequivocally the presence of shocks, nor is it
the highest found among planetary nebulae (Perinotto \& Corradi
\cite{Pe98}), it nevertheless supports such an interpretation.

Thirdly, significant deviations from the overall expansion pattern are
found in the kinematics of some knots, especially those located to the
west of the core and close to it. In that region, both the radial and the
tangential velocities are larger than in the rest of the nebula, and
the surface brightness is enhanced.

Finally, the determination of the distance deserves a special mention.
The main original aim of our multi-epoch imaging of Hen~2--147 was to compute
the expansion parallax of the nebula and derive its distance, a crucial
parameter (and often one of the most uncertain) in discussions of the
physics of any astrophysical object.  The expansion parallax is
traditionally considered as a robust, purely geometrical method to compute
distances, in which no assumptions need to be made about the physics of the
object. Hen~2--147 shows that the situation is far from simple; there are
clear indications (broadening of the emission lines, discrepancy with the
distance obtained via the Mira Period-Luminosity relation) that what we
measure in the plane of the sky is not the expansion of {\it matter}, but
rather the expansion of a {\it shock front}.  If we do not take this into
account, a straightforward application of the expansion parallax method
results in a distance of Hen~2--147 of 1.5$\pm$0.4~kpc.  If instead we
include the effects of shocks (see section 5.2.1), the distance to
Hen~2--147 scales up to 2.7$\pm$0.5~kpc, in agreement with the distance of
3.0$\pm$0.4~kpc obtained using the period-luminosity relation for the Mira.

A detailed spectroscopic analysis is essential to determine if shock
excitation is actually important in this nebula. Nevertheless, the large
difference in the distance determination, in the presence or absence of
shocks, shows that one must be extremely careful when estimating distances
via the expansion parallax method.  This has been demonstrated,
theoretically, by Mellema (\cite{Me04}) and Sch\" onberner et al.
(\cite{Sc05}) in the case of planetary nebulae.  The method is also widely
used to estimate distances for supernovae (Panagia \cite{Pa98}) and
classical novae (e.g. Chochol et al. \cite{Ch97}, Esenoglu et al.
\cite{Es00}, Krautter et al. \cite{Kr02}, Lyke et al. \cite{Ly03}), and the
same caution should be exercised in these studies.  We should also emphasize
that for a spatially resolved nebula, simultaneous study of the structure
and kinematics are essential to correctly determine the geometrical,
kinematical and projection parameters.

In the case of Hen~2--147, it is indeed the combination of a thorough
study of the nebular geometry and kinematics along the line of sight
and in the plane of the sky, together with the fact that an
independent distance estimation can be obtained from the near-IR data,
that makes our conclusion a robust one: {\it the two distance methods
agree if the expansion in the plane of the sky corresponds to the
expansion of a strong shock front}.  One possible test for the
presence of shocks would be to examine emission lines sensitive to
shocks (e.g.  the $\frac{H\alpha}{[N{\sc II}]654.8+658.3}$
vs. $\frac{H\alpha}{[S{\sc II}]671.7+673.1}$ diagnostic diagram, see
Hollis et al. \cite{Ho92}).  Note also that the distance obtained
using the Mira PL relation could be scaled down to the value of
1.7~kpc from the expansion parallax result, in the absence of shocks,
only if the (presumably circumstellar) extinction in $K$-band is 1.13
magnitudes larger than estimated in section 5.1 (corresponding to an
extinction larger by 10 magnitudes in $V$). This seems highly unlikely,
 although we cannot completely rule out this possibility.

For these reasons, we adopt a distance to Hen~2--147 of 2.9$\pm$0.4 kpc.
At this distance, the physical radius of the ring nebula is 0.04~pc. In
addition, the apparent expansion indicates that the nebula was ejected
some ~200 years ago. This is likely to be the epoch at which the white dwarf
companion of the Mira experienced a nova outburst, from which the system has
almost completely recovered (Munari \cite{Muna97}); the currently observed
spectrum of the core is characteristic of the quiescent phase of symbiotic
novae (Munari \& Zwitter \cite{MZ02}).

\begin{acknowledgements}

The authors wish to thank Vera Khozurina-Platais, Warren Hack, Anton
Koekemoer, and Richard Hook from the STScI for their kind help in
multidrizzle usage; Reinhard Hanuschik for the FLAMES data pipeline
processing; Mariano Santander, Corrado Giammanco, Bruce Balick and Vincent Icke for ideas and comments;
the following observers from SAAO who contributed to the near-infrared
photometry: Brian Carter, Greg Roberts, Johnathan Spencer Jones, Robin
Catchpole and Michael Feast; and the anonymous referee for pointing out an error in 
equation (1) in the first version of this paper.

\end{acknowledgements}

\onecolumn
\begin{longtable}{lcccc}
\caption[]{Hen~2--147: Near-Infrared Photometry}\\
\hline
\multicolumn{1}{c}{JD}   & $J$ & $H$ & $K$ & $L$ \\
\multicolumn{1}{c}{--2440000} &\multicolumn{4}{c}{(mag)}\\
\hline
\endfirsthead
\caption[]{Continued.}\\
\hline
\multicolumn{1}{c}{JD}   & $J$ & $H$ & $K$ & $L$ \\
\multicolumn{1}{c}{--2440000} &\multicolumn{4}{c}{(mag)}\\
\hline
\endhead
\noalign{\smallskip}
\hline
\endfoot
\noalign{\smallskip}
\hline
\endlastfoot
\noalign{\smallskip}
   4769.3& 6.88& 5.54& 4.80& 3.90\\
   4777.3& 6.94& 5.58& 4.85& 3.97\\
   4779.4& 6.95& 5.61& 4.88& 4.04\\
   4802.3& 6.93& 5.57& 4.86& 4.05\\
   4823.2& 6.74& 5.42& 4.74& 3.89\\
   5036.5& 7.51& 5.95& 5.11& 4.26\\
   5060.6& 7.44& 5.93& 5.09& 4.16\\
   5100.5& 7.20& 5.82& 5.05& 4.25\\
   5121.4& 7.14& 5.80& 5.04& 4.13\\
   5159.3& 7.20& 5.84& 5.10& 4.18\\
   5544.3& 6.82& 5.51& 4.82& 3.94\\
   5569.2& 6.54& 5.32& 4.68& 3.86\\
   5573.2& 6.50& 5.28& 4.64& 3.85\\
   5710.6& 6.69& 5.21& 4.51& 3.88\\
   5773.6& 7.16& 5.62& 4.84& 4.08\\
   5779.6& 7.20& 5.66& 4.87& 4.09\\
   5801.6& 7.14& 5.64& 4.86& 4.04\\
   5846.4& 6.78& 5.44& 4.73& 3.92\\
   5876.3& 6.67& 5.40& 4.73& 3.89\\
  6186.4*& 7.33& 5.82& 5.01& 4.21\\
   6199.4& 7.17& 5.68& 4.90& 4.09\\
   6219.5& 6.94& 5.55& 4.80& 3.99\\
   6225.4& 6.90& 5.55& 4.81& 3.99\\
   6266.3& 6.91& 5.62& 4.87& 4.01\\
   6493.6& 7.20& 5.63& 4.84& 4.16\\
   6541.5& 7.38& 5.76& 4.93& 4.15\\
   6548.5& 7.36& 5.76& 4.93& 4.19\\
   6569.5& 7.20& 5.70& 4.89& 4.10\\
   6596.4& 6.99& 5.59& 4.83& 4.05\\
   6640.3& 6.91& 5.57& 4.85& 4.02\\
   6704.2& 6.48& 5.22& 4.59& 3.86\\
   6825.6& 6.66& 5.16& 4.46& 3.87\\
   6848.6& 6.92& 5.37& 4.62& 3.92\\
   6867.5& 7.10& 5.51& 4.73& 4.01\\
   6895.6& 7.30& 5.66& 4.85& 4.14\\
   6897.6& 7.24& 5.66& 4.83& 4.12\\
   6897.6& 7.27& 5.65& 4.83& 4.09\\
   6923.5& 7.27& 5.65& 4.84& 4.07\\
   6961.4& 6.83& 5.43& 4.69& 3.86\\
   6985.3& 6.76& 5.43& 4.70& 3.87\\
   7010.3& 6.82& 5.52& 4.80& 3.88\\
   7073.2& 6.61& 5.31& 4.63& 3.83\\
   7235.6& 7.16& 5.58& 4.79& 4.06\\
   7280.5& 7.66& 6.01& 5.12& 4.29\\
   7328.4& 7.45& 5.85& 5.00& 4.50\\
   7358.3& 7.13& 5.69& 4.90& 4.11\\
   7384.3& 7.08& 5.68&     &     \\
   7581.6& 6.66& 5.19& 4.53& 3.88\\
   7609.5& 6.95& 5.42& 4.69& 4.00\\
   7634.6& 7.25& 5.65& 4.86& 4.15\\
   7669.5& 7.53& 5.88& 5.04& 4.29\\
   7688.5& 7.47& 5.86& 5.00& 4.22\\
   7705.4& 7.29& 5.75& 4.92& 4.13\\
   7742.4& 6.94& 5.55& 4.80& 3.98\\
   7782.3& 6.84& 5.51& 4.77& 3.96\\
   8030.5& 7.41& 5.86& 5.05& 4.26\\
   8067.4& 7.59& 6.04& 5.18& 4.31\\
   8089.3& 7.45& 5.96& 5.12& 4.27\\
   8121.2& 7.22& 5.87& 5.07& 4.18\\
   8142.3& 7.16& 5.82& 5.04& 4.20\\
   8388.5& 7.22& 5.64& 4.86& 4.18\\
   8427.4& 7.41& 5.79& 4.97& 4.23\\
   8457.3& 7.11& 5.64& 4.86& 4.07\\
   8813.4& 7.08& 5.60& 4.84& 4.12\\
   9168.3& 7.41& 5.86& 5.03& 4.24\\
   9236.3& 7.02& 5.66& 4.91& 4.14\\
   9445.6& 6.43& 5.02& 4.38& 3.78\\
   9503.5& 7.13& 5.61& 4.83& 4.09\\
   9583.3& 6.95& 5.61& 4.86& 4.05\\
  10212.6& 6.51& 5.09& 4.42& 3.79\\
  10256.4& 6.97& 5.48& 4.73& 4.06\\
  10500.6& 6.03& 4.81& 4.22& 3.57\\
  10589.5& 6.60& 5.15& 4.46& 3.85\\
  10618.4& 7.00& 5.39& 4.67& 4.00\\
  10717.2& 6.57& 5.31& 4.63& 3.85\\
  10949.6& 6.61& 5.18& 4.50& 3.86\\
  11056.6& 7.24& 5.78& 4.97& 4.12\\
  11234.6& 6.47& 5.14& 4.51& 3.81\\
  11298.5& 6.63& 5.18& 4.51& 3.87\\
  11329.4& 6.92& 5.41& 4.67& 3.96\\
  11388.3& 7.56& 5.90& 5.03& 4.27\\
  11417.3& 7.38& 5.82& 4.98& 4.14\\
  11599.7& 6.29& 4.98& 4.36& 3.63\\
  11674.5& 6.61& 5.15& 4.48& 3.84\\
  11685.5& 6.77& 5.25& 4.55& 3.95\\
  11709.5& 7.03& 5.48& 4.74& 4.06\\
  11743.4& 7.33& 5.73& 4.92& 4.22\\
  11781.3& 7.38& 5.84& 5.00& 4.14\\
  11931.6& 6.61& 5.33& 4.65& 3.83\\
  11963.6& 6.47& 5.12& 4.47& 3.76\\
  11976.6& 6.49& 5.10& 4.44& 3.75\\
  12033.5& 6.71& 5.20& 4.51& 3.87\\
  12070.4& 7.05& 5.46& 4.69& 3.99\\
  12091.5& 7.26& 5.62& 4.81& 4.09\\
  12116.4& 7.43& 5.78& 4.93& 4.16\\
  12154.3& 7.37& 5.78& 4.90& 4.08\\
  12181.2& 7.11& 5.63& 4.81& 4.02\\
  12324.6& 6.33& 5.02& 4.37& 3.62\\
  12383.5& 6.39& 4.97& 4.30& 3.61\\
  12429.4& 6.91& 5.39& 4.64& 3.94\\
  12534.3& 7.74& 6.14& 5.23& 4.29\\
  12691.6& 6.69& 5.38& 4.71& 3.92\\
  12753.6& 6.72& 5.26& 4.59& 3.95\\
  12801.5& 7.20& 5.58& 4.83& 4.15\\
  12823.4& 7.44& 5.77& 4.93& 4.24\\
  12857.3& 7.85& 6.09& 5.17& 4.37\\
  13062.6& 6.60& 5.25& 4.55& 3.75\\
  13118.6& 6.61& 5.16& 4.47& 3.74\\
  13140.5& 6.71& 5.23& 4.50& 3.74\\
  13177.4& 7.08& 5.48& 4.70& 3.99\\
  13193.4& 7.25& 5.61& 4.80& 4.06\\
  13203.4& 7.34& 5.71& 4.86& 4.03\\
  13228.3& 7.64& 5.94& 5.03& 4.23\\
  13262.2& 7.68& 6.02& 5.09& 4.16\\
  13415.6& 6.60& 5.28& 4.57& 3.62\\
  13452.7& 6.36& 5.02& 4.35& 3.56\\
  13511.5& 6.55& 5.10& 4.41& 3.82\\
  13562.3& 7.05& 5.48& 4.72& 4.03\\
\label{jhkl}
\end{longtable}
\twocolumn

\end{document}